\documentclass{article}
\usepackage[preprint]{tackling_climate_workshop_style}
\usepackage{geometry}
\usepackage{graphicx}
\usepackage{qtree}
\usepackage{booktabs}
\usepackage{siunitx}
\usepackage{ragged2e}
\usepackage{rotating}
\usepackage{adjustbox}
\usepackage{caption}
\usepackage[T1]{fontenc}
\usepackage{pdflscape}
\usepackage{nameref}
\usepackage{pdfpages}
\usepackage{float}
\usepackage{tikz}
\usepackage{tabularx} % Required for 'tabularx' environment

\usetikzlibrary{shapes.geometric, positioning, arrows.meta}
\newcolumntype{R}[1]{>{\RaggedRight}p{#1}}
\usepackage{amssymb}
\usepackage{xcolor}
\usepackage{verbatim}
\usepackage{subcaption}
\usepackage{array}
\usepackage{natbib}
\usepackage{hyperref}
\usepackage{mathtools}
\usepackage{xspace}
\usepackage{cleveref}
\usepackage{svg}
\newcommand*{\eg}{e.g.\@\xspace}

\usepackage{transparent}
\usepackage[toc,title,page]{appendix}
\definecolor{orangefull}{RGB}{230, 159, 0}
\colorlet{orange}{orangefull!20!white}
\definecolor{bluefull}{RGB}{86, 180, 233}
\colorlet{blue}{bluefull!20!white}
\definecolor{green}{RGB}{0, 158, 115}
\colorlet{green}{green!20!white}
\title{Turbine location-aware multi-decadal wind power predictions for Germany using CMIP6}
\author{%
  Nina Effenberger\thanks{corresponding author} \\
  Cluster of Excellence Machine Learning\\
  University of Tübingen\\
  \texttt{nina.effenberger@uni-tuebingen.de} \\
  \And
  Nicole Ludwig \\
  Cluster of Excellence Machine Learning\\
  University of Tübingen\\
  \texttt{nicole.ludwig@uni-tuebingen.de}}
\begin{document}

\maketitle

\begin{abstract}
    Climate change will impact wind and therefore wind power generation with largely unknown effect and magnitude. Climate models can provide insights and should be used for long-term power planning. In this work we use Gaussian processes to predict power output given wind speeds from a global climate model and compare the aggregated predictions to actual power generation. Analyzing past climate model data supports the use of CMIP6 climate model data for multi-decadal wind power predictions and highlights the importance of being location-aware. Our predictions up to 2050 reveal only minor changes in yearly wind power generation. We find that wind power projections of the two in-between climate scenarios SSP2-4.5 and SSP3-7.0 closely align with actual wind power generation between 2015 and 2023. Our analysis also reveals larger uncertainty associated with Germany's coastal areas in the North as compared to Germany's South, motivating wind power expansion in regions where future wind is likely more reliable. Overall, our results indicate that wind energy will likely remain a reliable energy source in the future.
\end{abstract}
%\ioptwocol

\section{Introduction}
\label{introduction}
To mitigate climate change, wind energy will play an essential role in future power supply \citep{barthelmie2021climate}. Efficient power planning should therefore account for natural wind variability as well as climate change by incorporating climate projections into multi-decadal predictions \citep[\eg][]{miao2023evaluation}. However, these climate projections have two main shortcomings: Their output resolutions are coarse due to the high (computational) complexity of climate models and are uncertain as they account for, among other things, unpredictable human behavior.

% people do downscaling
To overcome the issue of coarse spatial resolution (usually $\geq \SI{100}{\km}$) of general circulation models (GCMs), so-called downscaling techniques have been developed \citep[\eg][]{sun2024deep}. Downscaling, including statistical and machine learning methods \citep[\eg][]{langguth} and dynamical downscaling, can increase the spatial but also the temporal resolution of GCMs. For multi-decadal wind power predictions, where the primary goal is an accurate cumulative power prediction, \citet{effenberger2023mind} have shown that a temporal resolution of 6 hours is generally sufficient. An analogous observation has not yet been made for spatial resolutions; a high spatial resolution is often beneficial \citep[\eg][]{tamoffo2020process} and can resolve more physical processes and weather phenomena \citep{letson2020wrf} but requires careful selection \citep{pryor2020climate}.
% Data issue: for CMCIP6 there are no downscaled models yet -> we should use the models anyway 
For CMIP6 \citep{eyring2016overview}, the latest version of globally organized GCMs, no high-resolution regional model runs are available yet in contrast to its predecessor CMIP5, compare \citet{jacob2014euro}. To overcome this issue, \citet{bartok2019climate} have developed a climate projection dataset tailored for the European energy sector based on CMIP5. However, previous research revealed that CMIP6 and CMIP5 show differences in future wind resource projections for Europe \citep{carvalho2021wind} and CMIP6 showed better capability in simulating surface wind speeds across the entire Northern Hemisphere \citep{miao2023evaluation}. A critical point of climate models is their ability to integrate radiative forcing and represent different scenarios. In this context, \citet{jung2022review} show that the unlikely worst-case climate model scenario SSP5-8.5 \citep{hausfather2020emissions} is over-represented in current research. Therefore, while the plausibility of different scenarios is unclear \citep[\eg][]{pielke2022plausible}, there is a need for projecting realistic scenarios of CMIP6 for multi-decadal power prediction. 

% related work and their results regarding Germany
Several studies investigate potential changes in wind power resources due to climate change. The studies mostly differ in the data used and the study region considered. We refer to \citet{jung2022review} for an overview of recent studies on wind resource projections under climate change and summarize some main points and more recent work here. \citet{gernaat2021climate} investigate data from CMIP5 and find that changes in wind energy are uncertain with complex patterns across climate models; \citet{barkanov2024evolution} investigate raw CMIP6 data and reveal changes in European offshore renewable energy resources. \citet{martinez2024global} find a significant decline in wind resources by 2100 in CMIP6, particularly evident in the mid-latitudes of the Northern Hemisphere; for Germany they find negligible changes in wind power generation in the long-term future (2091–2100) under the high emission climate change scenario SSP5-8.5. Investigating CORDEX climate model data (compare \citep{jacob2014euro}) for 2025–2049, \citet{sander2021greenhouse} support this claim and find that climate change will affect wind energy in Germany only marginally. Several studies investigate regions out of the scope of this study \citep[\eg][]{nabipour2020modeling, martinez2022climate, he2023mapping}; all reveal similar results in terms of the complexity of spatial and temporal patterns. 

% difficulty of power forecasting
As most of the renewable power data is confidential, it is common to use wind speeds \citep{jung2020introducing} or wind speeds cubed \citep{miao2023evaluation} as a proxy for wind power. Most of the reviewed work considers gridded climate data only, however some research also incorporate turbine locations for more realistic power predictions \citep[\eg][]{tobin2016climate, jung2020introducing}. In this work, we further expand the framework of location-awareness by not only predicting turbine location-aware multi-decadal wind power but also validating these predictions with actual wind power generation. 

Using CMIP6 data directly, we account for the latest climate model updates. The framework of Gaussian processes (GPs) allows to additionally include turbine locations into our power projections and we show that these are similar to ground truth aggregated power generation. GPs have proven useful in recent wind power assessment studies \eg by \citet{moradian2024enhancing} or \citet{esnaola2024future} as well as downscaling climate variables \citep[\eg][]{chau2021deconditional, kupilik2024bias}. In most cases downscaling refers to increasing the resolution of gridded data (compare \cite{sun2024deep}) and one main advantage of GPs compared to other statistical downscaling approaches is that they do not rely on a grid. This makes them a natural choice for turbine location-specific downscaling. Additionally their probabilistic framework can be useful in the context of climate modeling where projections are usually associated with high uncertainty \citep{lehner2020partitioning}. 

Our analysis reveals that for reliable  multi-decadal wind power predictions taking turbine locations into account is even more important than the choice of climate scenario. To assess future wind power generation and identify promising areas for wind power expansion political decision makers should therefore rely on turbine location-aware projections. We do not only present a new approach for validating multi-decadal wind power predictions but also provide turbine location-aware predictions for Germany up to 2050. We describe our approach in \Cref{methods}, our results in \Cref{results} and discuss and conclude in \Cref{discussion} and \Cref{conclusion}.

\begin{comment}
    \textcolor{red}{info stuff}
But \cite{carvalho2021wind} show that “CMIP6 future wind resource projections for Europe show relevant differences when compared to CMIP5”. This is also pointed out by \cite{miao2023evaluation} who show that “ CMIP6 models show better capability in simulating surface wind speed over the whole Northern Hemisphere in terms of pattern correlation coefficient, standard deviation ratio, and root-mean-square difference than the CMIP5 models; and the biases compared with observations reduce apparently in Europe and Asia. Despite underestimating the magnitude of the trend, all CMIP6 models reproduce decreasing trends in wind speed, while nearly half of the CMIP5 models show rising trends opposite to the observations.” Miao et al. use wind power density, which equals wind speed cubed. However, this is not necessarily realistic. And there is a review by \cite{jung2022review}:  the most plausible scenarios are SSP2-4.5, SSP460, and SSP3-7.0. The worst-case scenario SSP5-8.5 is considered highly unlikely [18].
    
Increase resolution of ERA 5 using DEM data \cite{hu2023downscaling}, categorize real data into three categories depending on how well the real data aligns with ERA5. 

Uncertainty: "Over land, the majority of projection uncertainties is dominated by model uncertainty, followed by the internal variability and scenario uncertainty" \cite{zhang2024quantify}. 
Relating CMIP6 GCM climate data to wind power generation over Germany allows us to make wind power forecasts using the latest version of climate models. 
\end{comment}

\section{Methods}
\label{methods}
Our general approach includes 1) estimating wind speeds at turbine locations 2) extrapolating wind speeds to hub-height and 3) predicting the corresponding power output. We compare the wind speeds at turbine locations to predictions that do not consider actual turbine placement, but are instead based on gridded weather or climate datasets. We perform the same steps on these datasets but 1) use the wind speeds at grid points 2) extrapolate wind speeds to the average hub-height and 3) compute the power output using the most common turbine across the dataset. 

\subsection*{Data}
For our evaluation, we consider the gridded reanalysis dataset {ERA5} \citep{hersbach2020era5} and the gridded climate dataset \emph{MPI-ESM1.2-HR} \citep{muller2018higher} from CMIP6. Furthermore, we compare our predictions generated using these weather and climate datasets to aggregated transmission level power generation. The power data was collected from individual transmission system operators (TSOs) across Germany \citep{ods2024} and data provided by the German federal agency "Bundesnetzagentur" through the {SMARD} database \citep{bundesnetz}. To access the turbine locations and other static turbine data, we use a turbine dataset provided by \citet{manske_2023_8188601}. For the gridded data that covers Germany we set the boundaries in ERA5 to longitudes $\in [5°, 15°]$ and latitudes $\in [47°, 56°]$. In the CMIP6 model runs the boundaries of the box considered are longitudes $\in [5.63°, 15.0°]$ and latitudes $\in [47.22°, 55.63°]$ and use all climate scenarios available for \emph{MPI-ESM1.2-HR}, namely SSP1-2.6, SSP2-4.5, SSP3-7.0, SSP5-8.5. As suggested by \citet{effenberger2023mind} we use 6-hourly wind speed data. 

%\begin{figure}[H]
    \begin{figure}[h!]
    \begin{minipage}[b]{0.49\textwidth}
\centering
        \raisebox{-2em}{\includegraphics[width=\linewidth]{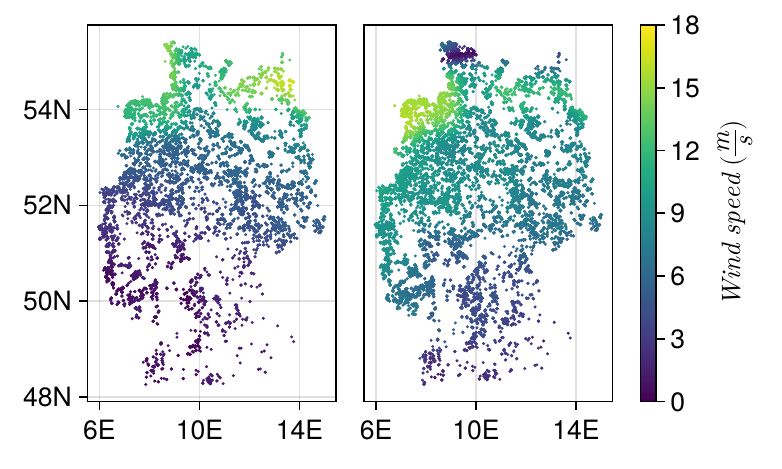}}
\caption{Turbine locations and the corresponding wind speeds on January 1st 2011 (left) and 2023 (right) respectively. There are more turbines in the North than the South and wind speeds are usually higher in the North.}
\label{fig:example-forecast}
\end{minipage}
\hspace{0.02\textwidth}
    \begin{minipage}[b]{0.49\textwidth}
\centering
        \resizebox{\columnwidth}{!}{%
\begin{tikzpicture}[scale=0.7]
    % Define colors
    \definecolor{lightblue}{HTML}{440154FF}
    \definecolor{darkblue}{HTML}{FFFFFF}

    % Timeline
    \draw[thick] (0,0) -- (12,0);

    % Draw the first block (2011 to 2023)
    \draw[fill=lightblue, draw=darkblue, thick] (0,-0.5) rectangle (12,0.5);
    \node at (6, 0) {\textcolor{darkblue}{ERA5}};

    % Draw the second block (2011 to 2015 and 2015 to 2023)
    \draw[fill=lightblue, draw=darkblue, thick] (0,0.5) rectangle (4,1.5);
    \draw[fill=lightblue, draw=darkblue, thick] (4,0.5) rectangle (12,1.5);
    \node at (2, 1) {\textcolor{darkblue}{TSO data}};
    \node at (8, 1) {\textcolor{darkblue}{SMARD data}};

    % Draw the third block (2011 to 2015 with an arrow showing continuation)
    \draw[fill=lightblue, draw=darkblue, thick] (0,1.5) rectangle (4,2.5);
    \draw[fill=lightblue, draw=darkblue, thick] (4,1.5) rectangle (12,2.5);
    \node at (2, 2) {\textcolor{darkblue}{CMIP6 historical}};
    
    % Draw continuation arrow behind the box
    \draw[-{Triangle[width=18pt,length=10pt]}, line width=18pt, draw=lightblue](12.1,2) -- (13.5,2);
    \node at (12.65,2) {\textcolor{darkblue}{2100}};
    \node at (8, 2) {\textcolor{darkblue}{CMIP6 SSPs}};

    % Draw time axis below the lowest block
    \draw[thick] (0,-1) -- (12,-1);

    % Draw years/stamps on the time axis
    \foreach \x/\year in {0/2011, 4/2015, 12/2023}
    {
        \draw[thick, darkblue] (\x,-1.2) -- (\x,-0.8);
        \node[below] at (\x,-1.4) {\year};
    }
\end{tikzpicture}
}
\caption{We use weather (ERA5), climate (CMIP6 historical and SSPs) and power data (TSO and SMARD) between 2011 and 2023. Due to limited data availability not all datasets are temporally aligned.}
\label{fig:data}
\end{minipage}
\end{figure}

%\end{figure}

\subsection*{Estimate wind speeds at turbine locations}
We compute wind speeds at turbine locations using Gaussian processes (GPs). A GP is a collection of random variables where any finite subset follows a multivariate normal distribution. A GP is defined by a mean function $\mu(\cdot)$ and a covariance function $k(\cdot, \cdot)$ that is a positive definite kernel, see \Cref{matern}. We consider the case where the output of the climate models is noisy, i.e. the underlying function $y$ is corrupted by Gaussian noise and therefore 
\begin{equation}
\label{sigma}
    y = f(x)+ \epsilon \mathrm{\;where\; } \epsilon \sim \mathcal{N}(0, \sigma^2).
\end{equation}
We compute $\sigma^2$ as the variance over time of the two model runs that are available on the ESGF website \citep{esgf}. To keep extreme values of the individual model runs we do not use the mean of the model runs as input but only the first model run \texttt{r1i1p1f1}. In GP regression we put a GP prior on $f$ and compute the posterior given data $D={(x_i,y_i)}^n_{i=1}=: \{\mathbf{X}, y\}$. The posterior is also a GP and can be computed analytically. For further details, we refer to \citet{pml1Book}.

\paragraph{Kernel choice}
We use a Matérn kernel of order $\frac{3}{2}$, which for inputs $x, x'$ and metric $d(\cdot, \cdot)$ is given by
\begin{equation}
\label{matern}
    k(x, x') = \lambda^2 \big(1 + \frac{\sqrt{3} d(x, x')}{\ell} \big) \exp\big(- \frac{\sqrt{3} d(x, x')}{\ell} \big)\text{\,,}
\end{equation}
where $d$ is the Euclidean metric $d(x, x') = ||x - x'||_2$ and $\lambda$ and $\ell$ are hyperparameters.
We model the wind speed $w$ at one location and time point using a single-output GP. The original data consists of wind velocities $u$ and $v$ and we first compute the wind speed as
\begin{equation}
    w = \sqrt{u^2+v^2}\text{\,.}
\end{equation}
To predict wind speeds at turbine locations we condition on the gridded spatial dataset at the same time point. Our predictions for past data between 2011 to 2023 are then compared to ERA5 predictions as well as actual power generation. For the high-resolution reanalysis data set ERA5 we choose a multi-output GP that models the wind velocities. We present results with multi-output GPs on historical data  in \Cref{fig:old-forecast}. Since this approach did not improve the results noticeably we use single-output GPs that work with wind speeds for the future predictions to avoid unnecessary high runtime. We give an example of the predictions for turbine locations at two example time points in \Cref{fig:example-forecast}.

\paragraph{Hyperparameter optimization}
We optimize the hyperparameters $\boldsymbol{\theta} = \{\lambda, \ell\}$ of the Matérn kernel $K$ in \Cref{matern} by maximizing the marginal likelihood 
\begin{equation}
    p(\mathbf{y} \mid \mathbf{X}, \boldsymbol{\theta}) = \mathcal{N}(\mathbf{y} \mid 0, K+\sigma_n^2 I) \text{\,.}
\end{equation}
Hyperparameters are optimized on the historical data from 2011 using gradient descent \citep{nocedal1999numerical} on the log marginal likelihood
\begin{equation}
\log p(\mathbf{y} \mid \mathbf{X}, \boldsymbol{\theta}) = 
-\frac{1}{2} \mathbf{y}^T (K + \sigma_n^2 I)^{-1} \mathbf{y} 
- \frac{1}{2} \log \left| K + \sigma_n^2 I \right| 
- \frac{n}{2} \log 2\pi 
\end{equation}

where $I$ is the identity matrix and $n$ the number of data points. During inference time the hyperparameters are then fixed and set to the average value of the historical run of 2011. We give more information regarding the variability of the hyperparameters in \Cref{appendix} and visualize the results of the hyperparameter optimization in \Cref{fig:0004_parameters_optim}.

\paragraph{Spatial uncertainty}
We investigate the posterior marginal standard deviation of the of wind speeds at turbine locations and give an example in \Cref{fig:posterior-var}. To better account for the large differences in average wind speeds over land and sea we normalize the standard deviation with the average wind speed at the same location. 

\subsection*{Extrapolate wind speeds to hub height and compute power}
\begin{figure}
    \centering
    \includegraphics[width=0.5\textwidth]{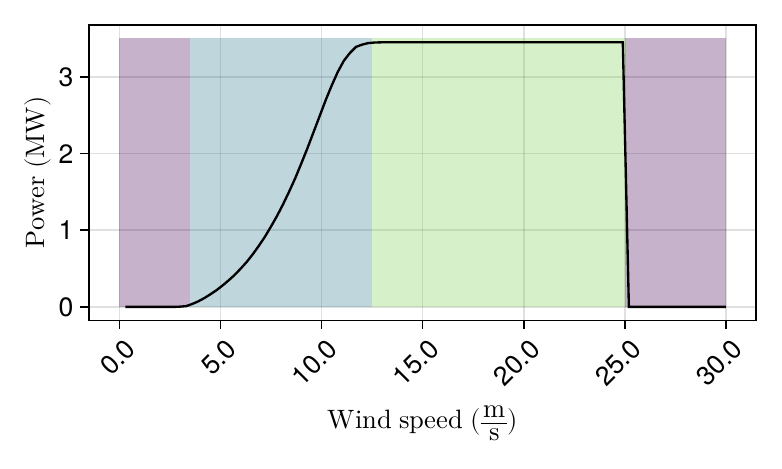}
    \caption{Turbine power curve of the Enercon E-53/800 turbine. No power is generated at very low and very high wind speeds (purple) and once the rated power has reached maximum power is generated in all cases (green). The relationship between wind speed and power output is almost cubic in the blue part.}
    \label{fig:power-curve}
\end{figure}

We predict wind speeds at turbine locations using GPs. Given wind speeds $w_{10}$ at a height of \SI{10}{\meter} \citep{cmip6-cop} the wind speed $w(z)$ at hub-height $z$ can be computed assuming a wind profile power law with 
\begin{equation}
    w(z)=w_{10}\cdot\left(\frac{z}{10}\right)^\alpha\text{\,.}
\end{equation}  
Following \citet{wan2019universal}, we set the wind shear coefficient to $\alpha=\frac{1}{7}$. We set the hub height for the gridded datasets to the mean (\SI{78.77}{\meter}) of the 2011 turbine dataset \citep{manske_2023_8188601}. 
To compute the wind power generation of each turbine, we feed the GP wind speed predictions at the turbine locations into turbine power curves, an example of such a curve is given in \Cref{fig:power-curve}. We choose a suitable power curve for each turbine we model by mapping the turbines from the python library windpowerlib \citep{sabine_haas_2024_10685057} to the static turbine data provided by \citet{manske_2023_8188601}. For each installed turbine in the German database, we choose the turbine in windpowerlib whose capacity is closest to the actual installed capacity. To model past yearly wind power generation, we account for all turbines installed in or before the respective year. For future wind power generation we account for all turbines in the database where the commission date is 2024 or earlier. For the gridded datasets, we choose the turbine that occurs most often (E-53/800), one of the smallest turbines in the database, and can not account for an increasing number of turbines as the power curve is applied to each grid point. The prediction of the total power generated at a time point $t$ is called $p_{\text{pred}}(t)$, which is the sum over all grid points or turbines. We perform linear bias correction by computing a factor $f$ that ensures that the cumulative power generation prediction after 365 days $P_{\text{pred}}(365\cdot4)$ equals the power $P_{\text{true}}(365\cdot4)$ that was generated in the considered year
\begin{equation}
    f = \frac{\sum_{i=1}^{365\cdot4}p_{\text{pred}}(i)}{\sum_{i=1}^{365\cdot4}p_{\text{true}}(i)} =: \frac{P_{\text{pred}}(365\cdot4)}{P_{\text{true}}(365\cdot4)}\text{\,.}
\end{equation}
This linear bias correction term should account for dispatch \citep[\eg][]{goransson2009dispatch} and other constant biases in wind power modeling. We correct the historical projections (2011 to 2014) with the true power generation of 2011, the past projections (2015 to 2023) with the true power generation of 2015 and future power generation with the wind power generated in 2023. 

\subsection*{Evaluation period from 2011 to 2023}
We compare historical runs of GCMs from 2011 to 2014 to ERA5 as well as actual power generation in Germany, furthermore we evaluate different CMIP6 scenarios between 2015 and 2023. An overview of the different datasets and how they temporally overlap can be found in \Cref{fig:data}. We compare the historical and scenario runs of CMIP6 to ERA5 as the latter is highly correlated with observational data  \citep{kaspar2020regional} and showed better performance in forecasting wind power generation than other reanalysis datasets such as MERRA2 in previous studies \citep{olauson2018era5}. As actual power generation is our true variable of interest, we compare the wind power predictions from the historical runs with the power generation reported by the four different German TSOs \citep{ods2024}. After 2015 aggregated wind power generation data of these TSOs is available on the SMARD database \cite{bundesnetz} which we compare to the power predictions from the climate scenario projections. 

\begin{figure}[H]
    \begin{minipage}[b]{0.49\textwidth}
        \centering
    \includegraphics[width=\linewidth]{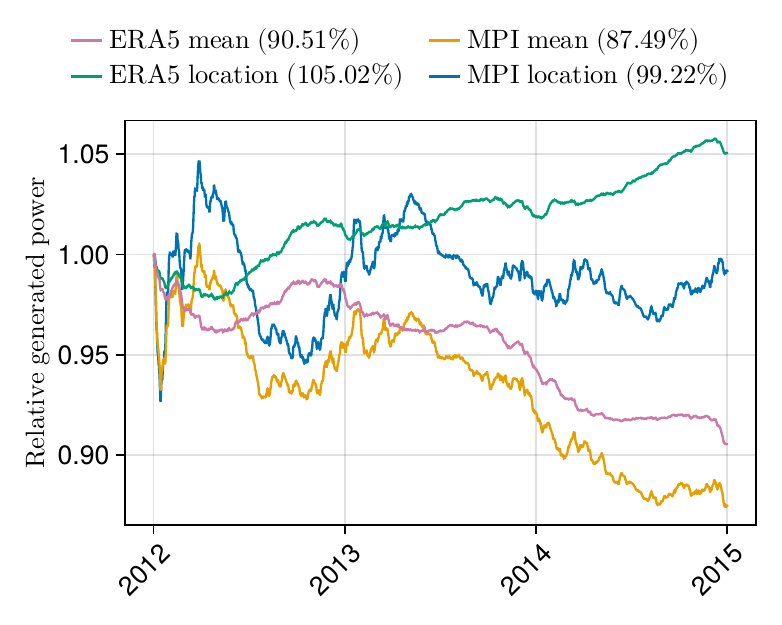}
    \caption{Power prediction using historical CMIP6 data and ERA5 relative to the true power generated. A value of 1.0 indicates a perfect prediction. It can be seen that location-aware predictions are closer to the actual power generated.}
    \label{fig:historical-forecast}
\end{minipage}
\hspace{0.01\textwidth}
    \begin{minipage}[b]{0.49\textwidth}
        \centering
    \includegraphics[width=\linewidth]{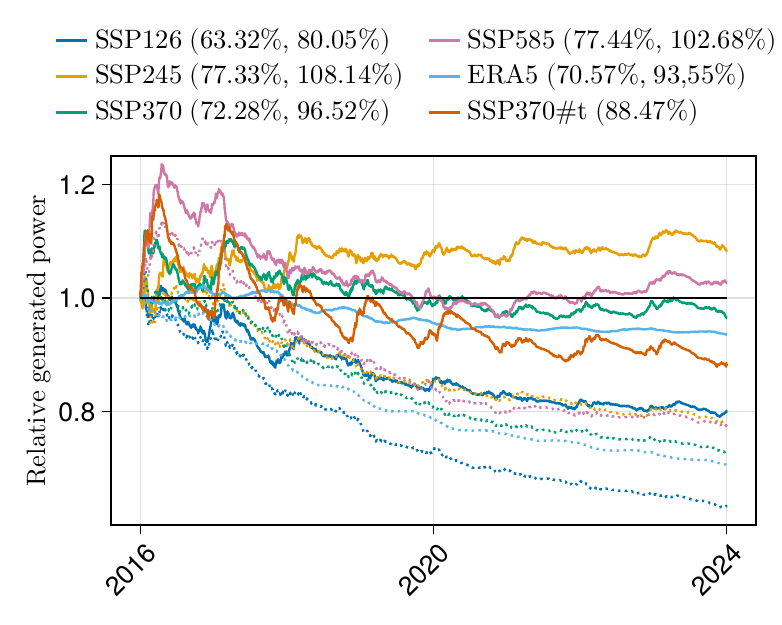}
        \caption{Power prediction using scenarios of one climate model relative to the actual power. The first number in brackets is the accuracy of the prediction without location (dotted lines), and the second is with location (solid lines).}
    \label{fig:power-forecast-cmip6}
    \end{minipage}
\end{figure}

\section{Results}
\label{results}
We divide our results into three parts: 1) validation of our method and investigation of past data, 2) future wind power projections and 3) spatial uncertainty quantification. Our results reveal that including turbine locations is very influential for multi-decadal wind power predictions. We further find that -- independent of the scenario considered -- the uncertainty of climate projections over Germany is higher in the coastal North than the mountainous South. 

\subsection*{Method validation}
Using GCM data and turbine locations, we predict wind power generation in Germany. For the historical period 2011 to 2014, location-aware cumulative power predictions with ERA5 overestimate wind power generation by 5.02\%, and the location-aware prediction using the historical run of the \emph{MPI-ESM1.2-HR} model considered underestimates power generation by 0.78\%, see \Cref{fig:historical-forecast}. In both cases, the accuracy of the non-location-aware prediction is lower, with an underestimation of 9.49\% and 12.51\% for ERA5 and CMIP6, respectively. In the future scenarios, we find that for our region and study period, SSP5-8.5--the worst-case reference scenario considered-- is closest (+2.68\%) to the true generated power if the prediction is location-aware, see \Cref{fig:power-forecast-cmip6}. However, the prediction that aligns closest with the turbine location-aware ERA5 prediction in terms of mean absolute error is the medium-to-high reference scenario SSP3-7.0 \citep{meinshausen2020shared}, see \Cref{appendix} and \Cref{tab:Mae}. If locations and with that, the increasing number of turbines are not considered, wind power generation is underestimated in the climate scenarios as well as ERA5. If the prediction is weighted by the number of turbines in a specific year (SSP3-7.0+\#t), wind power predictions get underestimated compared to the location-aware prediction (SSP3-7.0). 

\subsection*{Turbine location-aware multi-decadal wind power predictions in Germany using CMIP6}
\begin{figure}[H]
    \begin{minipage}[b]{0.49\textwidth}
    \centering
    \includegraphics[width=\linewidth]{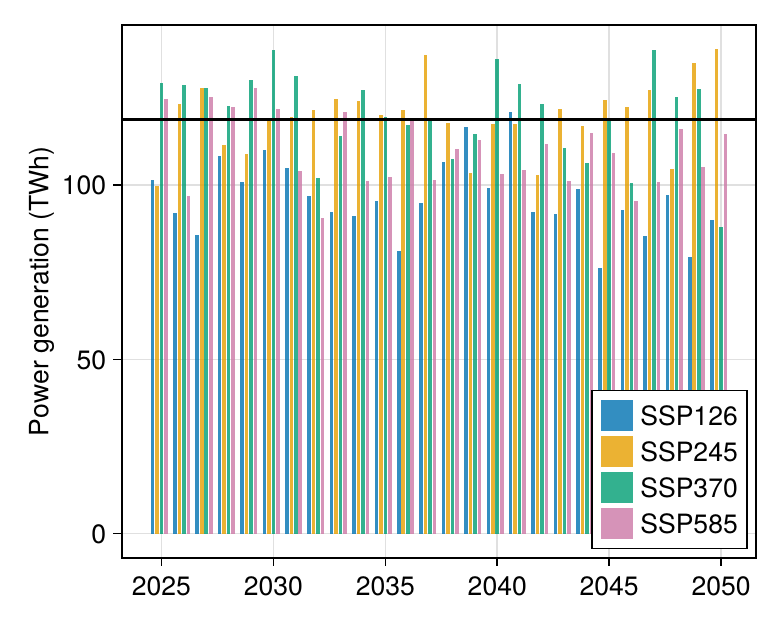}
    \caption{Yearly turbine location-aware power predictions for the different climate scenarios. The black line indicates the onshore wind power generation in 2023. On average wind power generation in SSP2-4.5 and 3-7.0 will be a bit higher than in 2023 while SSP1-2.6 and SSP5-8.5. project lower power generation.}
    \label{fig:prediction-cmip6}
    \end{minipage}
\hspace{0.01\textwidth}
    \begin{minipage}[b]{0.49\textwidth}
    \centering
    \includegraphics[width=\linewidth]{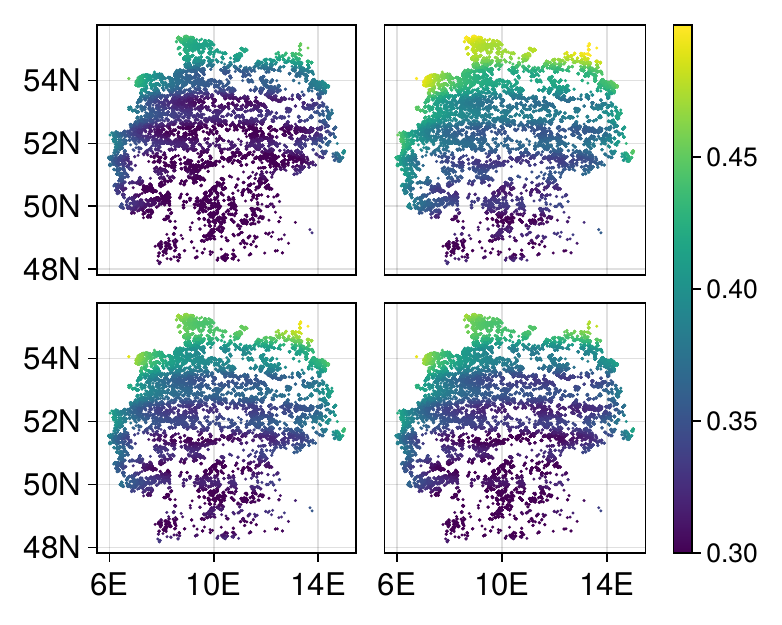}
    \caption{Average posterior standard deviation at the turbine locations in 2050 for SSP1-2.6 (top left), SSP2-4.5(top right), SSP3-7.0 (bottom left) and SSP5-8.5 (bottom right). The overall pattern is similar for all scenarios with higher uncertainties in the in the coastal North as compared to the mountainous South.}
    \label{fig:posterior-var}
    \end{minipage}
\end{figure}

We predict turbine location-aware wind power for Germany up to the year 2050 and compare location-aware to non-location-aware predictions. We present yearly results in \Cref{fig:prediction-cmip6} and show the cumulative predictions in \Cref{fig:cumulative-power-2050}. For 2050, location-aware predictions result in expected power generation between 87.87 TWh and 138.98 TWh, see \Cref{tab:forecasts}. For the scenarios SSP1-2.6, SSP3-7.0 and SSP5-8.5 being location-aware results in lower expected cumulative power between 2025 and 2050, namely by 14.77\%, 13.85\% and 84.34\% respectively. Only in the scenario SSP2-4.5 the cumulative location-aware power prediction is 1.34\% higher than the non-location-aware prediction.

In Germany in 2023 a total of 448,85 TWh of electric power was fed into the grid with a relative amount of 118,78 TWh (26.46\%) being wind power \citep{bundesnetz}. To contextualize the reported results we compare the power predictions for 2050 to the expected power consumption of 506 TWh in 2050 as reported by the \cite{energytarget}. The predictions of the different scenarios reveal an expected power generation between 87.87 TWh and 138.98 TWh which is between 17.37 and 27.47\% of the total power target of 2050. 

\begin{table}[h!]
    \centering
    \small
    \caption{Power generation predictions using the different climate scenarios pathways of the MPI-ESM1.2-HR. In the 2023 persistence prediction (last row) we do not correct for the extra day in leap years.}
    \begin{tabularx}{\textwidth}{|X|X|X|X|}
    \hline
        Pathway& Power prediction \newline2050 (in TWh)& Average prediction \newline 2045-2050 (in TWh) & Cumulative prediction \newline 2025-2050 (in TWh) \\ \hline
        SSP1-2.6 &  90.01&88.85&2501.88\\
        SSP2-4.5 & 138.98&125.54&3109.20\\
        SSP3-7.0 & 87.87&115.80&3134.75\\
        SSP5-8.5 & 114.65&106.22&2858.18 \\
        Wind power in 2023 & 118.78& 118.78 & 3088.34 \\
        \hline
    \end{tabularx}
    \label{tab:forecasts}
\end{table}

\subsection*{Uncertainty quantification}
The framework of GPs enables integrating the different ensemble members into the projection and to quantify the uncertainty. As only two runs per SSP that are not temporally aligned are available, the variance per timesteps of these is difficult to interpret. In our model setup we choose to use the variance of the two model runs per scenario as noise $\sigma$ (see \Cref{sigma}) which results in a spatially meaningful posterior variance, see \Cref{fig:posterior-var}. The results mainly reveal two insights: The normalized posterior standard deviation is higher for turbine locations closer to the coast in the North and varies more with latitude than longitude. This is in line with the hyperparameter optimization which resulted in larger values i.e. smoother functions of longitude compared to latitude.

\section{Discussion}
\label{discussion}
% This is a general discussion of results
Our results indicate that multi-decadal wind power predictions are possible with GCM output and turbine locations.
% location-awareness
In many experiments, non-location-aware predictions differed substantially from location-aware predictions, indicating that accounting for the number and locations of turbines is crucial. 
% which model is the best and what is the forecasst 
Our results investigating past data reveal that for the region and time considered, turbine-location-aware power predictions using SSP3-7.0 are most similar to the predictions with ERA5 data and to the ground-truth generated power. Investigating future climate projections further reveals that the differences between the two in-between scenarios SSP2-4.5 and SSP3-6.0 and wind power generation in 2023 are minor. This indicates that wind energy is likely to be a reliable power source in the future. In general, accounting for turbine locations resulted in a smaller spread of the four climate scenarios compared to non-location-aware predictions, indicating that climate change could have smaller impacts on wind power generation than expected when investigating raw climate model data. Due to these minor changes in expected power generation, our results are in line with other studies (\citep[\eg][]{sander2021greenhouse, martinez2022climate}) and underscore that wind power and storage expansion can likely compensate for the impacts of climate change. Our results regarding the spatial uncertainty of the projections further motivate wind power expansion in the South of Germany--despite the on average lower wind speeds--as less uncertain wind conditions are to be expected. Our results reveal that the best and worst case scenarios represent only the extremes and do not account for the full spectrum of possible outcomes. This underscores the controversial results \citep{schwalm2020reply} by \citet{jung2022review} and \citet{hausfather2020emissions}, i.e. the need to investigate all scenarios available and not only SSP5-8.5 and SSP1-2.6.

In the following, we will discuss some assumptions we made in this work. Most of these are a consequence of lacking data of curtailment and individual turbines. 
%bias correction
Our linear power generation bias correction can not account for potentially changing curtailment. Generally, the value of bias correction is unclear \citep{maraun2016bias}, and therefore, we decided against bias correcting wind speeds. Investigating past data reveals that the correlation between ERA5 and wind power generation is not constant, indicating changes in curtailment or technical improvement of newly installed turbines. While curtailment data is partially available \citep[e.g.][]{joos2018short} most of it is confidential. %Ground truth: 
The true power generation we use to validate our results is therefore non-optimal as it includes curtailment and other manual interventions. The predictions generated using the gridded reanalysis dataset ERA5 are also just a proxy for the actual wind power potential. 

% bias correction
An additional reason for not bias correcting wind speeds is the limited availability of hub-height wind data and power data in general (compare \cite{effenberger2022collection}), which often limits research in the field of renewable energy modeling. The wind data we use is \SI{10}{\meter} surface wind speed data, which we vertically extrapolate using a wind profile power law which has known shortcomings \citep{touma1977}.
%Location-specific vs location-aware
Due to the lack of data at hub-heights these extrapolated wind speeds can not be validated. Therefore, while the cumulative sum over all locations is valuable, we can also not validate the power predictions of individual turbines. This means that the predictions we make are only location-aware, they are not location-specific. While vertical extrapolation could, in general, be improved \citep[e.g][]{crippa2021temporal}, the high complexity of the atmosphere further complicates finding or learning a better parameterization of the vertical wind profile.
Another simplification we make during power prediction is using deterministic power curves. Given ground truth wind power generation data at individual turbine locations, one could, for example, learn probabilistic power curves as done in \citet{yun2021probabilistic}. %However, we do not have access to such ground truth location-specific power data, and we do not have access to the original power curves either. 
%and resulted in choosing a global climate model over a local one, extrapolating wind speeds to hub-height, computing power output using power curves and comparing forecasts to ERA5 and aggregated power generation. 
%Data availability is a big issue in power modeling in general. 
%RCMs
%\citet{joos2018short} show that curtailment in Germany in 2016 was $\sim 4.4\%$. \\
%Wind power computation:
Overall, access to wind speed observations at hub height, the corresponding power data and theoretical wind power curves could further improve such predictions.

%How do we continue -> i.e. work in progress
In future work, our set-up with GPs can be used to investigate other potential turbine scenarios in Germany which can help political decision makers in selecting turbine locations as well as expanding storage capacities. Furthermore, the ability of GPs to quantify uncertainty has not been fully exploited in this study; this could for example be improved by using large ensembles \cite[e.g.][]{olonscheck2023new} and taking more than two ensemble members into account and adjusting the noise-level $\sigma$ in \cref{sigma} accordingly. 
%Limited generalizability 
Future research should also emphasize on validating the methodology for larger regions which requires a lot of effort \citep[\eg][]{zhang2021global}, due to the lack of a common database for wind turbine installations. Additionally, our setup is promising for investigating physics-informed GP kernels \citep[\eg][]{pfortner2022physics}.
\section{Conclusion}
\label{conclusion}
Using Gaussian processes to investigate past data from historical as well as scenario climate model runs we find that multi-decadal wind power predictions using the MPI-ESM1.2-HR model is promising. We also show that accounting for turbine locations is important and results in more accurate predictions as compared to non-location-aware predictions. Our study demonstrates that while climate change may bring minor changes to wind power generation in Germany by 2050, wind energy will likely remain a reliable power source under most climate scenarios. Furthermore, the greater uncertainty in Northern coastal regions, compared to the South, emphasizes the importance of location-specific strategies to enhance wind power reliability in the upcoming years. 

\section*{Acknowledgements}
Funded by the Deutsche Forschungsgemeinschaft (DFG, German Research Foundation) under Germany’s Excellence Strategy – EXC number 2064/1 – Project number 390727645 and the Athene Grant of the University of Tübingen. The authors thank the International Max Planck Research School for Intelligent Systems (IMPRS-IS) for supporting Nina Effenberger. The authors thank Luca Schmidt and the anonymous reviewers of the Neurips 2024 Climate Change AI workshop for valuable feedback on earlier versions of the manuscript.

\bibliographystyle{plainnat}
\bibliography{bib}

\clearpage
\begin{appendices}
\counterwithin{figure}{section}
\counterwithin{table}{section}
\section{Supplementary Material}
\label{appendix}
\subsection{Hyperparameters}
\begin{figure}[h]
    \centering
    \includegraphics[width=0.5\linewidth]{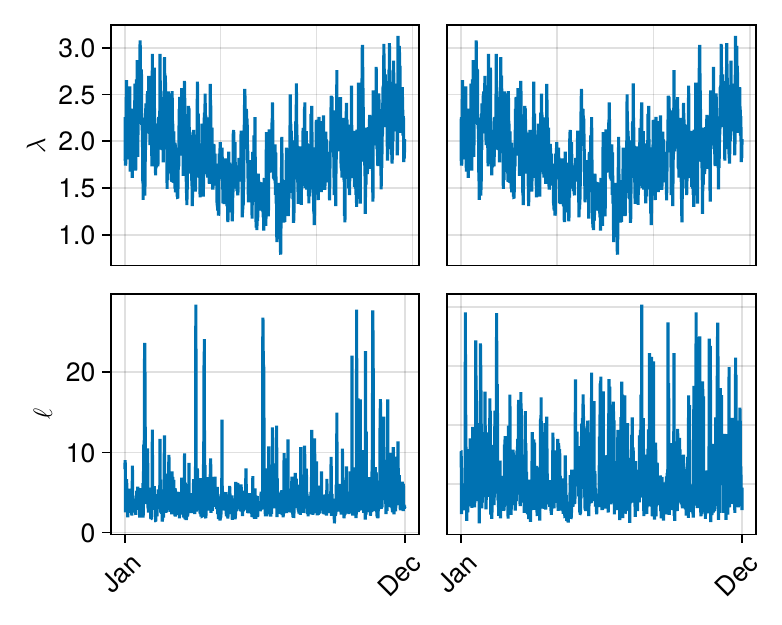}
    \caption{Results of the hyperparameter optimization. On average, $\ell$ is larger for the longitudes (left) than the latitudes. $\lambda$ expresses a seasonal pattern.}
    \label{fig:0004_parameters_optim}
\end{figure}

The length scale $\ell$ controls the width of the kernel, i.e. larger $\ell$ values result in smoother functions that are more correlated over larger distances. $\lambda$ controls the vertical scale of the kernel. Hyper parameter optimization for different time steps reveals that $\ell$ is more likely to vary for single time steps while $\lambda$ shows a seasonal pattern indicating on average less variability during the summer. The results for hyperparameter optimization is shown in \Cref{fig:0004_parameters_optim}. 

\begin{figure}[h!]
    \begin{minipage}[b]{0.49\textwidth}
        \centering
    \includegraphics[width=\linewidth]{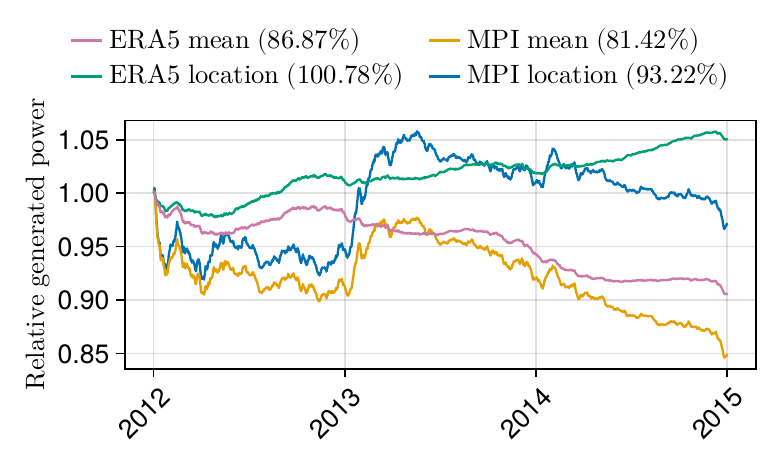}
    \caption{Power prediction using historical CMIP6 data and ERA5 relative to the true power generated using multi-output GPs. A value of 1.0 indicates a perfect prediction. It can be seen that location-aware predictions are closer to the true power generated.}
    \label{fig:historical-forecast}
\end{minipage}
\hspace{0.01\textwidth}
    \begin{minipage}[b]{0.49\textwidth}
        \centering
    \includegraphics[width=\linewidth]{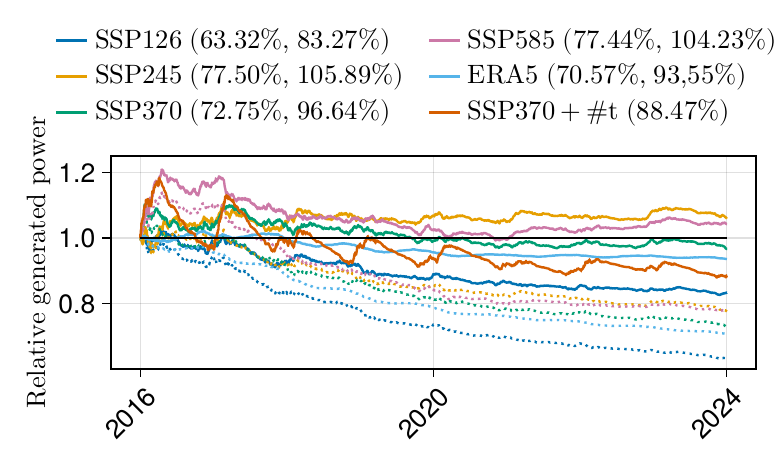}
        \caption{Power prediction relative to the true power generated using scenarios of one model using multi-output GPs. The first number in brackets is the accuracy of the prediction without location (dotted lines), and the second is with location (solid lines).}
    \label{fig:power-forecast-cmip6}
    \end{minipage}
\label{fig:old-forecast}
\end{figure}

\begin{figure}[H]
 \begin{minipage}[b]{0.49\textwidth}
        \centering
    \centering
\includegraphics[width=\linewidth]{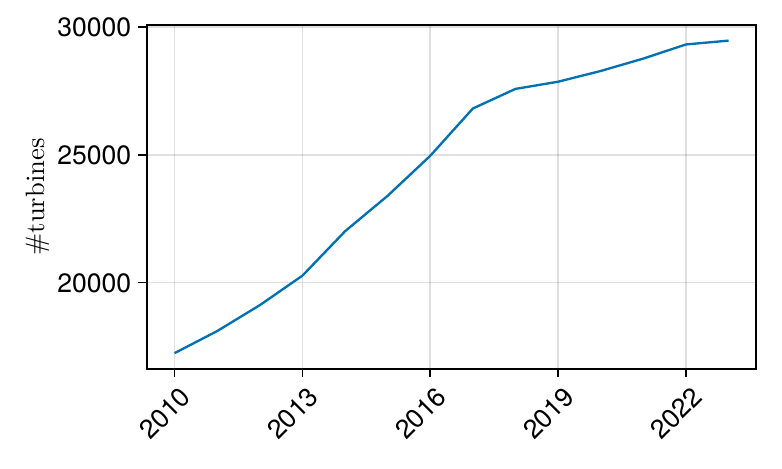}
    \caption{Number of wind turbines in Germany between 2010 and 2023. It can be seen that wind power is expanding.}
    \label{fig:num-turbines}
\end{minipage}
\hspace{0.01\textwidth}
    \begin{minipage}[b]{0.49\textwidth}
        \centering
    \includegraphics[width=\linewidth]{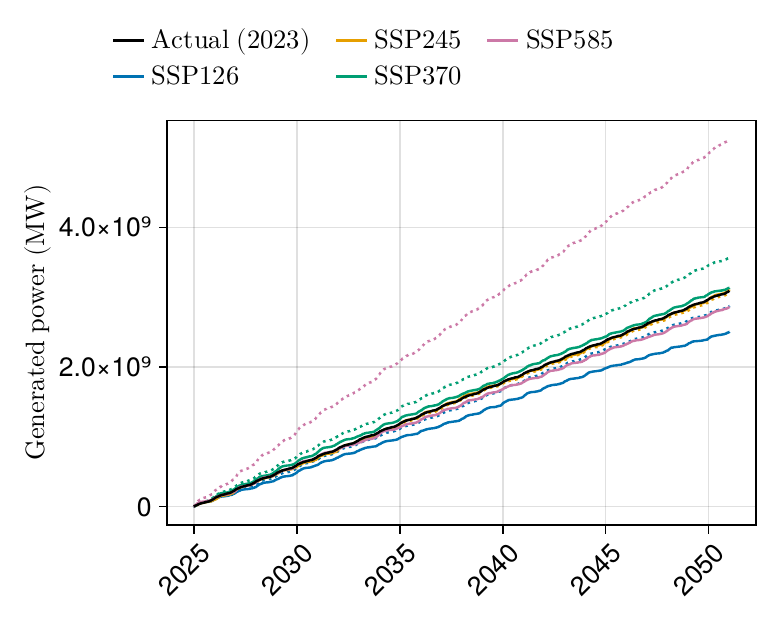}
    \caption{Cumulative power generation predictions using different climate model scenarios compared to the the power generation of 2023.}
    \label{fig:cumulative-power-2050}
    \end{minipage}
\end{figure}

To asses how similar the relative power predictions of the climate scenarios are compared to ERA5 we compute the mean absolute error between the ERA5 cumulative prediction $P_\text{ERA}$ and the four climate scenarios $P_\text{SSP}$ for all cumulative power predictions $n$ as
\begin{equation}
    MAE = \frac{\sum_{i=1}^n P_\text{ERA}(i)-P_\text{SSP}(i)}{n}
\end{equation}
\begin{table}[h]
    \centering
        \caption{MAEs of the relative cumulative power predictions of the climate scenarios compared to ERA5. SSP3-7.0 is closest to ERA5 in terms of MAE. }
    \begin{tabular}{|l|l|}
    \hline
    Climate scenario & MAE \\     \hline
         SSP1-2.6&0.10 \\
         SSP2-4.5&0.11 \\
         SSP3-7.0&0.04 \\
         SSP5-8.5&0.07 \\    \hline
    \end{tabular}
    \label{tab:Mae}
\end{table}
\end{appendices}
\end{document}